\documentclass[12pt,preprint]{aastex}
\newcommand{\lsim}{\, \lower2truept\hbox{${< \atop\hbox{\raise4truept\hbox{$\sim$}}}$}\,}
\newcommand{\gsim}{\, \lower2truept\hbox{${> \atop\hbox{\raise4truept\hbox{$\sim$}}}$}\,}

\newcommand{\puncspace}{\ifmmode\,\else{\ifcat.\C{\if.\C\else\if,\C\else\if?\C\else%
\if:\C\else\if;\C\else\if-\C\else\if)\C\else\if/\C\else\if]\C\else\if'\C%
\else\space\fi\fi\fi\fi\fi\fi\fi\fi\fi\fi}%
\else\if\empty\C\else\if\space\C\else\space\fi\fi\fi}\fi}
\newcommand{\SP}{\let\\=\empty\futurelet\C\puncspace}

\shorttitle{UV emission of Stephan's Quintet}
\shortauthors{Xu et al.}

\begin{document}

\title{UV emission and Star Formation in Stephan's Quintet}

\author{C. Kevin Xu \altaffilmark{1}, 
Jorge Iglesias-P\'{a}ramo\altaffilmark{2}, 
Denis Burgarella\altaffilmark{2},  
R. Michael Rich\altaffilmark{3}, 
Susan G. Neff\altaffilmark{4},
Sebastien Lauger\altaffilmark{2}, 
Tom A. Barlow\altaffilmark{1},
Luciana Bianchi\altaffilmark{5},
Yong-Ik Byun\altaffilmark{6}, 
Karl Forster\altaffilmark{1},
Peter G. Friedman\altaffilmark{1},
Timothy M. Heckman\altaffilmark{7},
Patrick N. Jelinsky\altaffilmark{8},
Young-Wook Lee\altaffilmark{6},
Barry F. Madore\altaffilmark{9,10},
Roger F. Malina\altaffilmark{2},
D. Christopher Martin\altaffilmark{1},
Bruno Milliard\altaffilmark{2},
Patrick Morrissey\altaffilmark{1},
David Schiminovich\altaffilmark{1},
Oswald H. W. Siegmund\altaffilmark{8},
Todd Small\altaffilmark{1},
Alex S. Szalay\altaffilmark{7},
Barry Y. Welsh\altaffilmark{8}, and
Ted K. Wyder\altaffilmark{1}}

\altaffiltext{1}{California Institute of Technology, MC 405-47, 1200 East
California Boulevard, Pasadena, CA 91125}
\altaffiltext{2}{Laboratoire d'Astrophysique de Marseille, BP 8, Traverse
du Siphon, 13376 Marseille Cedex 12, France}
\altaffiltext{3}{Department of Physics and Astronomy, University of
California, Los Angeles, CA 90095}
\altaffiltext{4}{Laboratory for Astronomy and Solar Physics, NASA Goddard
Space Flight Center, Greenbelt, MD 20771}
\altaffiltext{5}{Center for Astrophysical Sciences, The Johns Hopkins
University, 3400 N. Charles St., Baltimore, MD 21218}
\altaffiltext{6}{Center for Space Astrophysics, Yonsei University, Seoul
120-749, Korea}
\altaffiltext{7}{Department of Physics and Astronomy, The Johns Hopkins
University, Homewood Campus, Baltimore, MD 21218}
\altaffiltext{8}{Space Sciences Laboratory, University of California at
Berkeley, 601 Campbell Hall, Berkeley, CA 94720}
\altaffiltext{9}{Observatories of the Carnegie Institution of Washington,
813 Santa Barbara St., Pasadena, CA 91101}
\altaffiltext{10}{NASA/IPAC Extragalactic Database, California Institute
of Technology, Mail Code 100-22, 770 S. Wilson Ave., Pasadena, CA 91125}

\begin{abstract}
We present the first {\sl GALEX} UV images of the well known
interacting group of galaxies, Stephan's Quintet (SQ).  We detect
widespread UV emission throughout the group. However, there is no
consistent coincidence between UV structure and emission in the
optical, $H\alpha$, or HI. Excluding the foreground galaxy
NGC~7320 (Sd), most of the UV emission 
is found in regions associated with the two spiral members
of the group, NGC~7319 and NGC~7318b, and the
intragroup medium starburst SQ-A. The
extinction corrected UV data are analyzed to investigate the overall
star formation activity in SQ.  It is found that the total star
formation rate (SFR) of SQ is 6.69$\pm 0.65$ M$_\sun$ yr$^{-1}$. Among
this, 1.34$\pm 0.16$ M$_\sun$ yr$^{-1}$ is due to SQ-A. This is in
excellent agreement with that derived from extinction corrected
H$\alpha$ luminosity of SQ-A.  The SFR in regions related to NGC~7319
is 1.98$\pm 0.58$ M$_\sun$ yr$^{-1}$, most of which (68\%) is
contributed by the disk. The contribution from the 'young tail' is
only 15\%. In the UV, the 'young tail' is more extended ($\sim 100$ kpc)
and shows a loop-like structure,
including the optical tail, the extragalactic HII regions
recently discovered in H$\alpha$, and other UV emission regions
discovered for the first time.  The UV and optical colors of the
'old tail' are consistent with a single stellar population of age $t
\simeq 10^{8.5 \pm 0.4}$ yrs.  The UV emission associated with
NGC~7318b is found in a very large ($\sim 80$ kpc) disk, with a net
SFR of 3.37$\pm 0.25$ M$_\sun$ yr$^{-1}$.  Several large UV emission
regions are 30 --- 40 kpc away from the nucleus of NGC~7318b. 
Although both NGC~7319 and
NGC~7318b show peculiar UV morphology, their SFR 
is consistent with that of normal Sbc galaxies, indicating
that the strength of 
star formation activity is not enhenced by interactions.

\end{abstract}


\section{Introduction}
Stephan's Quintet (hereafter SQ) 
includes NGC~7317 (E), binary galaxies NGC~7318a (E) and NGC~7318b
(Sbc pec), Sy2 galaxy NGC~7319 (Sbc pec), and the foreground galaxy
NGC~7320 (Sd). Two very long parallel optical tidal tails ($> 100$
kpc; Arp \& Kormendy 1972) extend from the south end
of NGC~7319 toward another galaxy NGC~7320c on the east of SQ.  The HI
observations (Shostak et al. 1984; Williams et al. 2002, hereafter
W02) show HI tails following the optical tails. A large scale shock
front ($\sim$40 kpc) in the intragroup medium (IGM) between NGC~7319
and NGC~7318b was first discovered by Allen \& Hartsuiker (1972) as a
radio emission ridge, then confirmed by high resolution X-ray maps
(Trinchieri et al. 2003).  Moles et al. (1997, hereafter M97)
suggested a 'two-intruders' scenario for the history of SQ: an old
intruder (NGC~7320c) stripped most of the gas from group members, and
a new intruder (NGC~7318b) is currently colliding with this gas and
triggered the large scale shock. There are still many uncertainties
about this scenario.  These include the question whether both
optical tails, one of age $\sim 10^8$ yrs ('young tail') and another
of age $\sim 5\; 10^8$ --- $10^9$ yrs ('old tail'), are triggered
by two passages of the same galaxy NGC~7320c over NGC~7319 (M97).
Also, based on new HI maps which show that the HI in the region
occupied by NGC~7318b is clearly separated into two clumps with
distinctively different velocities, W02 challenged the conclusion
that NGC~7318b was {\it not} affected by interaction
until its collision with SQ starting about $10^7$ yrs ago (S01).
\begin{figure*}
\plotone{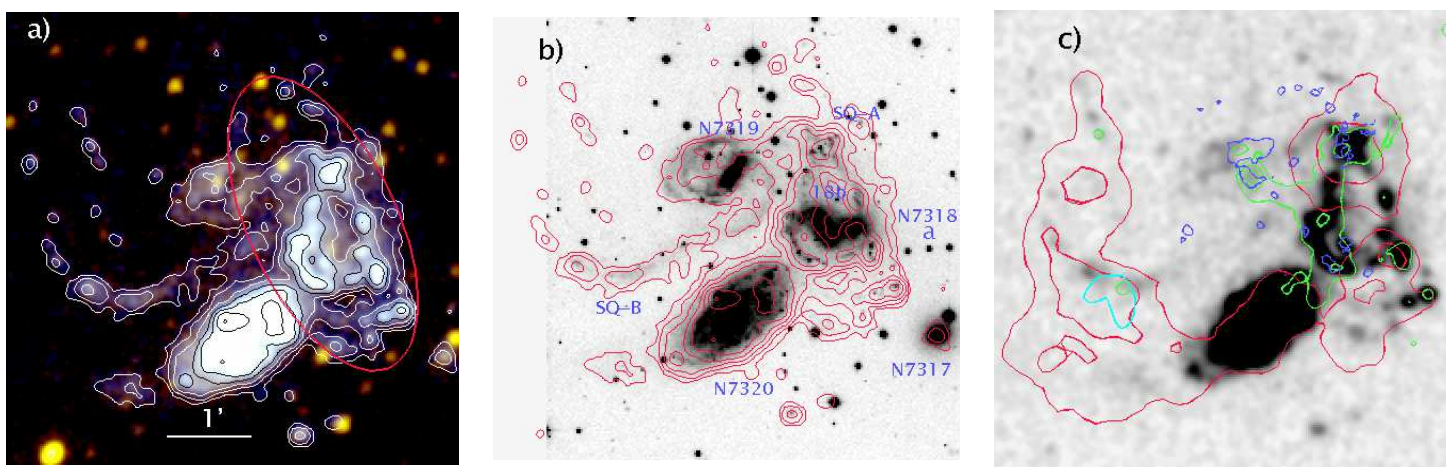}
\vskip-0.3cm
\caption{ {\bf (a)} Composite UV color image of SQ (blue: FUV, green:
FUV+NUV, red: NUV). In order to match the NUV resolution, the FUV
image is smoothed to a beam of FWHM = 7$''$.3, and r.m.s noise of the
smoothed map is 29.40 mag arcsec$^{-2}$.  The overlaid FUV contour
levels are 28.20, 27.33, 26.69, 25.94 and 25.19 mag arcsec$^{-2}$.
North is up and east on the left.  The scale of 1 arcmin is given,
corresponding to 27 kpc if the distance of SQ is assumed to be 94~Mpc
(H$_0$=70 Mpc$^{-1}$ km$^{-1}$~s). The red ellipse outlines the UV
disk associated with NGC~7318b.  {\bf (b)} The same FUV contours
overlaid on an B-band image.  {\bf (c)} The FUV image compared with the
H$\alpha$ (green contours, X99), HI (red contours, W02), 
CO from BIMA (blue contours, Gao \& Xu 2000) and CO from
IRAM (cyan contours, Lisenfeld et al. 2003).  }
\vskip-0.8cm
\end{figure*}

The star formation activity in SQ is apparently very much
influenced by the interactions.  The most spectacular star formation
region in SQ is the IGM starburst SQ-A, which is associated with a
bright MIR (15$\mu m$) source (Xu et al. 1999, hearafter X99) just
beyond the northern tip of the shock front. It is triggered by the
same NGC~7318b/IGM collision that triggered the shock (Xu et al. 2003,
hearafter X03).  The 'young tail', the brighter one among the two
optical tails, is also active in star formation. Hunsberger et
al. (1996) found 13 'tidal dwarf galaxy candidates' along this tail.
A bright HII region (SQ-B) is detect in both H$\alpha$ (Arp 1973) and
MIR (X99). Recently, Mendes de Oliveira et al.  (2004, hereafter
MdO04) discovered four intergalactic HII regions in a region north of
the 'young tail', possibly associated with the stripped ISM of
NGC~7319 (S01). The H$\alpha$ observations (Arp 1973; Vilchez \&
Iglesias-P\'aramo 1998; Plana et al. 1999; S01) reveal numerous huge
HII regions along several arms of NGC~7318b. Hunsberger et al. (1996),
Iglesias-P\'aramo \& Vilchez (2001), and Mendes de Oliveira et
al. (2001) classified them as tidal dwarf galaxy candidates, though
S01 argued that the crossing time of NGC~7318b ($\sim 10^7$ yrs) is
too short for any tidal effects. There has been no estimate of
the total star formation rate in SQ in the literature. 
Therefore it is not clear whether
the overall star formation activity is enhanced by the interactions.
Also, because a clear picture for
the interaction history of SQ is still missing, most results in the literature
linking the star formation activity to interactions 
in SQ were at best suggestive.

In this letter we present the first UV images of SQ obtained using
GALEX (Martin et al. 2004). The GALEX observations are sensitive to
very low levels of star formation averaged over $\sim 10^8$
yrs, the time scale of the tidal effects. This enables the first
quantitative study on the overall star formation and its distribution
in SQ. Comparisons between the UV, optical, HI, and the emission in
other wavebands put new constraints on the star formation
history and on the relation between star formation and galaxy-galaxy
interaction.

\section{GALEX Observations}
SQ was observed by GALEX on 2003-08-23, 2003-09-05 and 2003-09-06 in 5
orbits with total exposure time of 3327s. Data
reduction was done using standard GALEX pipeline (IR0.2 calibration,
Morrissey et al. 2004). The r.m.s noise
is 27.65 mag arcsec$^{-2}$ and 28.11 mag
arcsec$^{-2}$ in the FUV (1530{\AA}) and NUV (2310{\AA}), respectively. The UV
magnitudes are in the AB system. The FWHM of the FUV beam is 4$''$.8,
and that of the NUV beam 7$''$.3.
\begin{figure*}
\plotone{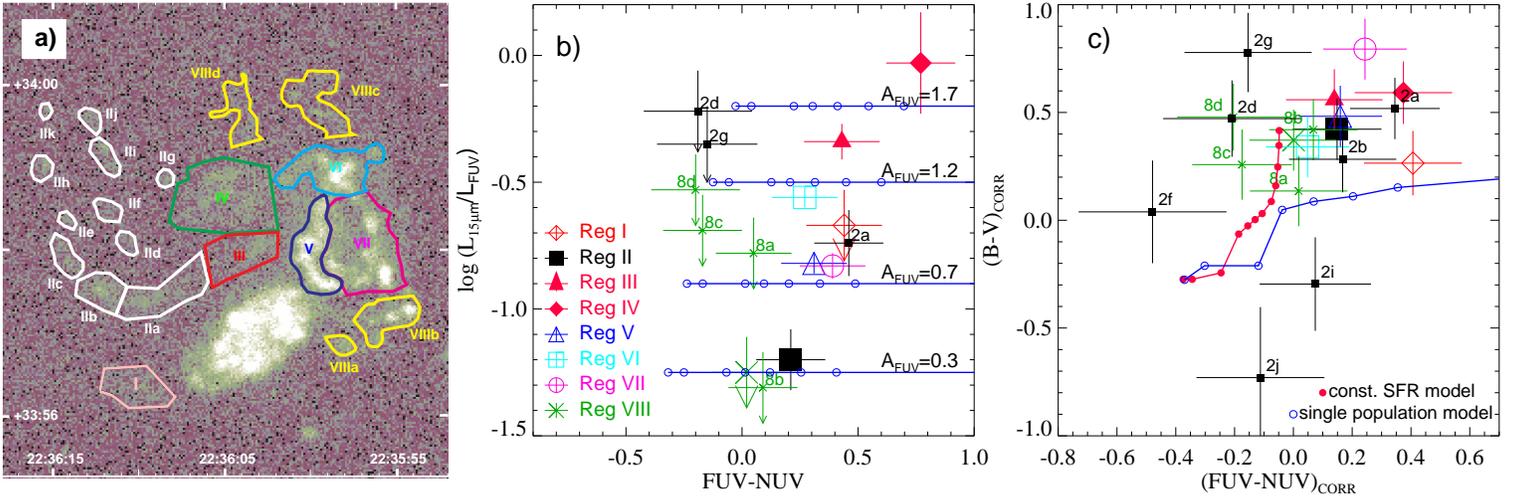}
\caption{ {\bf a)} Definition of regions listed in Table~1.
The boundaries are chosen according to FUV contours.
{\bf b)} L$_{15\mu m}$/L$_{FUV}$ v.s. FUV-NUV color plot.
The FUV and NUV data are corrected only for the foreground
Galactic extinction. The arrows indicate upper limits of
the 15$\mu m$ flux. The blue open circles linked by the blue
lines are the single population models (PEGASE burst models)
reddened by
the corresponding A$_{FUV}$. From left to right, the models
are with age t = 1, 2, 5, 20, 50, 100, 200, 500, 1000 Myr.
{\bf c)} B-V v.s. FUV-NUV color-color plot. Both colors are
corrected for the Galactic and the internal extinction.
The blue circles linked with blue line are the same
single population model (without reddening) as plotted in
Fig.2b. The red filled circles linked by red line are
PEGASE models assuming constant SFR with ages of 
t = 1, 2, 5, 20, 50, 100, 200, 500, 1000, 2000, 5000, 10000
Myr.}
\vskip-0.5cm
\end{figure*}

The GALEX image (Fig.1a) looks very different from the optical image
(Fig.1b). The foreground Sd galaxy NGC~7320 is the most prominent UV
source, but as 
NGC~7320 is not a member of SQ we do not discuss it further.  A
striking feature of the UV emission is that there are large regions
that do not coincide with significant optical light from the galaxies
or optically detected tidal features.
In the merger which gives rise to the Antennae (Hibbard et al. 2004)
the UV emission is seen to follow the tidal tails and merger body closely,
while being of $\sim 10^8\rm yr$, of age simlar to the distributed
UV emission in SQ.

There is
an extended UV disk (outlined by the red ellipse in the UV image)
centered between the nuclei of NGC~7318a and NGC~7318b.  It is most
likely to be associated with NGC~7318b and the contribution from the
NGC~7318a is insignificant.  This is because NGC~7318a has the optical
apparence of an ordinary elliptical galaxy and no HII region or HI gas
in this area is likely to be associated with it (M97;
S01).  The UV disk, with a physical size of $\sim 80$
kpc, is about 1$'$ larger (along the major axis) than the disk found
by S01 on the B-R image. The emission peaks in the south side of the UV
disk are in good agreement with those in the HI gas in this
region which has the same redshift of NGC~7318b (W02).
This agrees with the results of Thilker et al. (2004) who
found that the UV emission regions in outer disk of M83
are typically associated with HI structures.
The north tip of the disk goes beyond the HI gas boundary (Fig.1c).
This discrepancy between the UV and the cold gas may well
be due to the shock. SQ-A is also located within this disk.

The Sy2 nucleus of NGC~7319 is faint in the UV, indicating very high
extinction within the AGN,
consistent with the Sy2 classification.
Both the 'young tail' and the 'old tail' are detected.
The 'young tail' looks more like a 'loop' in the UV image.  Compared
to the optical tail, it extends further in the northeast direction
and includes all the 4 intergalactic HII regions found by MdO04.  
The UV emission within the NGC~7319 disk
shows a disturbed morphology. It is interesting to
note that no HI gas has been detected in NGC~7319 disk (Fig.1c; W02). 
The UV emission in the northeast
part of the disk is along two strong optical arms. It {\it does not}
coincide with the H$\alpha$ emission, nor with 
the CO emission (Fig.1c).
The elliptical galaxy
NGC~7317 is faint in the UV, and there is no indication that it has
been involved in any recent interactions with other members of SQ.

\section{Star Formation in SQ} 
The UV data are analyzed quantitatively to study the
star formation activity in different regions in SQ. The results
are presented in Table~1. For both the FUV and NUV
magnitudes, a calibration uncertainty of 0.1 mag is
assumed (Morrissey et al. 2004). The FUV magnitude, the
FUV-NUV color and the B-V color in Table~1 are corrected
for the foreground Galactic extinction (E(B-V)=0.079 and
the extinction curve of Cardelli et al. 1989). For regions
detected in ISOCAM 15$\mu m$ map (X99),
the internal extinction is determined using the 
luminosity ratio $\log(L_{15\mu m}/L_{FUV})$
according to the following formulae:
\begin{equation}
Y=\log(L_{IR}/L_{FUV})=\log(L_{15\mu m}/L_{FUV})+1.04,
\end{equation}
\begin{equation}
A_{FUV} =-0.033\; Y^3 + 0.352\; Y^2 + 1.196\; Y+ 0.497.
\end{equation}
The relation $\log(L_{IR})=\log(L_{15\mu m})+1.04$ is taken from Chary
\& Elbaz (2001). Eq.~(2) is taken from Buat et al. (2004).
Other regions are either undetected or outside the
ISOCAM 15$\mu m$ map. For them the internal extinction is estimated
using the average HI column density (W02) and the
dust-to-gas ratio of the solar neighborhood ($\tau_B=5.8\; 10^{-22}$
atoms$^{-1}$ cm$^{2}$, Savage and Mathis 1979). The foreground screen
model and a modified Calzetti attenuation curve (Buat et al. 2004)
are assumed. For the
extinction estimated using the HI, the error is assumed to be 100\%.
Estimates of the internal extinction of Reg.~VI (SQ-A) using the
MIR-to-FUV ratio and using the HI column density, respectively, are
found to be consistent with each other.  Star formation rate is
estimated using the extinction corrected FUV luminosity according to
the formula of Kennicutt (1998).  The age of the stellar population
responsible for the UV emission is determined using the extinction
corrected FUV-NUV color and the single population synthesis model
(i.e. the burst model) of Fioc \& Rocca-Volmerange (1997,
PEGASE).
 This is compared with the age of the stellar population
responsible for the optical emission, determined using the extinction
corrected B-V color and the same synthesis model.

\noindent{\bf SFR of SQ-A (Reg.~VI)}: 
The SFR of 1.34$\pm 0.16$ M$_\sun$ yr$^{-1}$
estimated from the extinction corrected FUV luminosity is
in excellent agreement with that
from the extinction corrected H$\alpha$ luminosity
(1.45 M$_\sun$ yr$^{-1}$) by X03. On the other hand,
the extinction estimated from the MIR-to-FUV ratio,
A$_{FUV} = 0.76\pm 0.06$, is significantly lower
than that indicated by the Balmer decrement
(A$_{H\alpha}\sim$ 0.6 -- 2). This suggests that
O stars and B stars which are responsible for the ionizing and
non-ionizing UV radiation, respectively,
have different distributions relative to the
dust. The extinction
corrected B-V color suggests an underlying
stellar population of $\sim 500$ Myr,
consistent with the results of Gallagher et al. (2001).

\noindent{\bf Star formation in regions related to NGC~7319}: 
These include Reg.~I, II, III, and IV. Among the net SFR
of 1.98$\pm 0.58$ M$_\sun$ yr$^{-1}$,
68\% is contributed by the disk (Reg. IV), and 15\%
from the 'young tail' (Reg. II).
There is a disk/tidal-tail gradient among the undereddened
UV colors of Reg.~IV, III
and II (Fig.2b). However,
the extinction corrected UV colors are consistent
with each other (Fig.2c), indicating that the color gradient is mainly
due to the dust reddening, and there is no significant difference
in the age of the UV population in the disk and in the tail.
The extinction corrected UV and optical colors of Reg. I (old tail) 
are consistent with a single population of age $t \simeq 10^{8.5 \pm 
0.4}$ yrs, somewhat younger than that estimated by S01 
from dynamic arguments (6--12 $\times 10^8$ yrs).

Reg.~II includes 11 subregions.
IIa, IIb and IIc correspond to the optical tail. The age of the UV
population in IIa and IIb is $\sim 10^8$ yrs while that of the optical
population is $\sim 5$ $10^8$ yrs, indicating an old underlying
population possibly stripped from the disk of NGC~7319 
in the encounter. 
The bright H$\alpha$/MIR source SQ-B is in the east end of IIa,
signaling active current star formation.  IIc has a much younger UV
population ($\sim 10^7$ yrs). It is outside our B and V image.  
The extragalactic HII regions found by MdO04 are also in Reg.~II. 
IIi (c and d in MdO04) shows very young ages in both UV
color and optical color ($\sim 10^6$ yrs), consistent with the
EW(H$\alpha$) result of MdO04. IIj, which is north of IIi and is not
in the list of MdO04, shows similar properties as IIi.  The UV
population of IIe (b in Md004) has an age of $\log(t)= 7.7+/-0.7$,
older than that estimated by MdO04 (5.6 Myr) from EW(H$\alpha$).  In
the UV image these subregions, together with other subregions which
are discovered for the first time, form a loop-like structure. Given
that many of these subregions were also detected in other bands
(e.g. optical and H$\alpha$) and available redshift data (S01; MdO04)
are consistent with they all having the same redshift, this
structure is very likely real.  M97 suggested that the young
tail is triggered by a high velocity ($\sim 700$ km/sec) passage of
NGC~7320c through NGC~7319 1.5 $10^8$ yrs ago. However, the
recently measured redshift of NGC~7320c is almost identical to that of
NGC~7319, indicating instead a slow passage (S01). Therefore in
order for NGC~7320c to move to its current position, the
NGC~7319/N7320c encounter must have occured $\ga 5\; 10^8$ yrs ago.
This is close to the age of the old tail, but older than that of the
young tail (M97). An alternative scenario is that the young tail is
triggered by a close encounter between NGC~7319 and the elliptical galaxy
NGC~7318a. Since NGC~7318a is about 3 times closer to NGC~7319 than
NGC~7320c, the time argument is in favor of this new scenario. Also
the 'UV loop', on the other side of NGC~7319 with regard to NGC~7318a,
looks very similar to the 'counter tidal loop' found frequently in the
dynamic simulations of galaxy-galaxy interactions (see, e.g., Fig.2 of
Toomre \& Toomre 1972, t=2 10$^8$ yr).

\noindent{\bf Star formation in regions related to NGC~7318b}: 
These include Reg.~V, VII, and VIII. The peaks of the 
UV emission in Reg.~V correspond to a chain of
HII regions in one of star forming arms of NGC~7318b
(S01; Gallagher et al. 2001). 
Therefore most of the UV emission is likely associated with
the star formation rather than with the shock.
The total SFR is 3.37$\pm 0.25$ M$_\sun$ yr$^{-1}$.
This is very close to the SFR of the Milky Way ($\sim 4\pm 2$
M$_\sun$ yr$^{-1}$, Mezger 1988), a normal Sbc galaxy.
Reg.~VIII includes 4 subregions, VIIIa and VIIIb
are in the south end of the UV disk, and VIIIc and 
VIIId in the north end. They are 30 --- 40 kpc
away from the nucleus of NGC~7318b. These 
are even larger than the largest distance among the UV emission regions
in the extreme outer disk of M83, studied by Thilker et al. (2004).
Whether the remarkable size of this UV disk, the peculiar morphology
of the HI gas, and the long and open arms (tidal tails?) 
of NGC~7318b are
due to a head-on collision with NGC~7318a about 10$^8$ yrs
ago (X03), is an interesting subject for future studies.
The UV and optical colors
of VIIIa are consistent with a single population of
$\sim 10^8$ yrs. 

\noindent{\bf Summary remarks}: Every time when SQ is observed in a
waveband for the first time, it reveals new, spectecular phenomona
related to its complex interaction history. The long term fate of the
star forming regions in the IGM starburst SQ-A, in the remarkably
extended 'UV loop' associated with the 'young tail', and in the very
large UV disk of NGC~7318b is of great interest. This can have
far-reaching inference to the interaction induced star formation in
multi-galaxy systems which may play important roles in the early
universe.

\noindent{\it Acknowledgments}:
GALEX (Galaxy Evolution Explorer) is a NASA Small Explorer, launched
in April 2003.  We gratefully acknowledge NASA's support for
construction, operation, and science analysis for the GALEX mission,
developed in cooperation with the Centre National d'Etudes Spatiales
of France and the Korean Ministry of Science and Technology.


\begin{deluxetable}{llllllllll}
\tablewidth{0pt}
\tablecaption{Star Formation Properties in SQ Regions}
\tablehead{
\colhead{(1)} & \colhead{(2)} & 
\colhead{(3)} & \colhead{(4)} & \colhead{(5)} &
\colhead{(6)} & \colhead{(7)} &
\colhead{(8)} & \colhead{(9)} &
\colhead{(10)} \\
\colhead{Name} & \colhead{Reg.} & 
\colhead{FUV} & \colhead{log(L$_{FUV}$)} & \colhead{SFR} 
& \colhead{FUV-NUV} & \colhead{log($t_{UV}$)} 
& \colhead{B-V} & \colhead{log($t_{BV}$)} 
& \colhead{A$_{FUV}$} \\ 
\colhead{} & \colhead{} & 
\colhead{mag} & \colhead{L$_\sun$} & \colhead{M$_\sun$ yr$^{-1}$} &
\colhead{mag} & \colhead{yr} &
\colhead{mag} & \colhead{yr} &
\colhead{mag} 
}
\startdata
Old Tail & I & 19.74$\pm 0.12$ &  8.49$\pm 0.08$ & 0.060$\pm 0.013$ & 0.44$\pm 0.16$ & 8.4$\pm0.2$ & 0.28$\pm 0.15$ & 8.5$\pm 0.4$    & 0.17$\pm 0.17$   \\
Young Tail & II & 18.14$\pm 0.11$ &  9.18$\pm 0.07$  & 0.296$\pm 0.051$ & 0.21$\pm 0.15$ & 7.9$\pm 0.3$ & 0.45$\pm 0.14$ & 8.7$\pm 0.3$  & 0.31$\pm 0.14$  \\
Tail/Disk Overlap & III & 19.34$\pm 0.12$ & 9.17$\pm 0.06$   & 0.285$\pm 0.046$ & 0.43$\pm 0.16$ & 7.9$\pm 0.3$ & 0.69$\pm 0.14$ & 8.9$\pm 0.3$  & 1.47$\pm 0.11$   \\
NGC~7319 Disk\tablenotemark{a} & IV & 18.19$\pm 0.11$ & 9.84$\pm 0.15$ & 1.341$\pm 0.571$ & 0.77$\pm 0.15$ & 8.3$\pm 0.3$ & 0.77$\pm 0.14$ & 9.0$\pm 0.3$ & 2.00$\pm 0.37$    \\
Shock Front & V & 16.87$\pm 0.10$ & 9.87$\pm 0.05$ & 1.450$\pm 0.164$  & 0.31$\pm 0.14$ & 7.9$\pm 0.3$   & 0.55$\pm 0.14$ & 8.8$\pm 0.2$   & 0.76$\pm 0.06$  \\
SQ-A & VI & 17.32$\pm 0.10$ & 9.84$\pm 0.05$ & 1.335$\pm 0.156$  & 0.27$\pm 0.14$ &  7.4$\pm 0.6$  & 0.44$\pm 0.14$ & 8.7$\pm 0.3$ & 1.12$\pm 0.07$   \\
NGC~7318b Inner Disk & VII & 16.76$\pm 0.10$ & 9.91$\pm 0.05$ & 1.561$\pm 0.179$ & 0.39$\pm 0.14$ & 8.1$\pm 0.3$ & 0.86$\pm 0.14$ & 9.3$\pm 0.3$  & 0.74$\pm 0.06$   \\
NGC~7318b Outer Disk & VIII & 17.72$\pm 0.11$ & 9.33$\pm 0.06$ & 0.360$\pm 0.052$ & 0.02$\pm 0.15$ & 7.4$\pm 0.5$ & 0.38$\pm 0.14$ & 8.7$\pm0.2$ &  0.10$\pm 0.10$ \\
total& & 15.36$\pm 0.10$ & 10.54$\pm 0.05$  & 6.687$\pm 0.646$ & 0.33$\pm 0.14$ & ... & ...  & ... & ...   \\
... &  & &  &  &  &    &  &    \\
\enddata
\tablecomments{
Table 1 is published in its entirety in the electronic edition
of The Astrophysical Journal Letters. Column (3) --- FUV corrected for
the foreground Galactic extinction. Column (4) --- extinction
corrected FUV luminosity (corrected
for both the Galactic and the internal extinction). The error includes
the uncertain of the extinction correction. Column (6) ---
UV color corrected only for the Galactic extinction. Column (7) ---
stellar population age derived from extinction corrected UV color,
using the single population model. The error includes
the uncertain of the extinction correction.
Column (8) --- optical color corrected only for the Galactic extinction.
Column (9) --- stellar population age derived from extinction 
corrected optical color, using the single population model. 
}
\tablenotetext{a}{Contribution from the AGN is subtracted from
the UV, optical and MIR flux densities. 
}
\end{deluxetable}

\end{document}